# New insight into quantifying vacancy distribution in self-ion irradiated tungsten: a combined experimental and computational study

*Zhiwei Hu [a,1], Jintong Wu [b,1], François Jomard [c], Fredric Granberg [b], Marie-France Barthe [a, *]*

[a] CEMHTI, CNRS, UPR3079, University of Orléans, F-45071 Orléans, France

[b] Department of Physics, Post-office box 43, FIN-00014 University of Helsinki, Finland

[c] Groupe d'Etude de la Matière Condensée, CNRS, UVSQ, 45 avenue des Etats-Unis, 78035 Versailles cedex, France

## Abstract

In this work, we propose a new approach based on positron annihilation spectroscopy to estimate the concentration of vacancy-type defects induced by self-ion irradiation in tungsten at room temperature, 500, and 700°C. Using experimental and Two-component density functional theory calculated annihilation characteristics of various vacancy clusters $V_n$ (n=1-65) and a positron trapping model associated with the simulated annealing algorithm, vacancy cluster concentration distribution could be extracted from experimental data. The method was validated against simulation results for room-temperature irradiation and transmission electron microscopy observations for higher temperatures. After irradiation at 500 and 700 °C, small clusters (<20 vacancies, ~0.85 nm) undetectable by TEM were unveiled, with concentrations exceeding $10^{25}$ m$^{-3}$, significantly higher than the concentration of TEM-visible defects ($10^{24}$ m$^{-3}$). Moreover, incorporating an oxygen-vacancy complex is deemed necessary to accurately replicate experimental data in samples subjected to high-temperature irradiation.

Tungsten (W) and other refractory metals are promising candidates for the first wall and divertor in future fusion power plants. It will be subjected to neutron irradiation and plasma exposure, both of which can cause macroscopic degradation, such as swelling [1] or bubbles [2,3] and/or fuzz formation on the surface [4–6]. To better understand the origin of these degradations, it is essential to investigate the underlying mechanisms. Combining experiments with modeling is a highly useful approach achieving this goal. It is therefore essential to carry out a quantitative analysis of irradiation-induced microstructure evolution, which may be influenced by the strong interactions between vacancy or interstitial defects and light element impurities (LEs) [7–9].

Positron annihilation spectroscopy (PAS) is a widely employed non-destructive technique for characterizing atomic-scale vacancy defects, particularly in metallic materials due to its exceptional sensitivity [10,11]. Positron, the antiparticle of the electron, possesses properties nearly identical to those of an electron, except for its positive charge. Upon entering a material, positrons lose their kinetic energy until they are thermalized and diffuse through the interatomic spaces. Open volumes, such as vacancy defects, act as efficient traps for positrons due to Coulomb forces. In materials with a low defect concentration, the positron diffusion length can exceed one hundred nanometers [12], before the positron annihilates with a nearby electron, a valence, or a core one in a delocalized state of perfect *Lattice*. Some positrons may even diffuse back to the sample surface, providing additional information about surface annihilation characteristics and effective diffusion length, $L_{eff}^+$. However, in materials with a high concentration of vacancy defects, the diffusion of positrons is limited due to the efficient trapping by defects. In such cases, the $L_{eff}^+$ can typically be reduced to only a few tens of nanometers (see supplementary III a). Positron annihilation is currently implemented in two main types of spectroscopy: Doppler Broadening (DB) spectroscopy and Positron Annihilation Lifetime Spectroscopy (PALS). These techniques measure two distinct annihilation characteristics: (i) the momentum distribution of annihilated electron-positron pairs (or DB spectrum). (ii) the positron lifetime, which is the time between the entrance of the positron in solid and its annihilation with an electron. From the momentum distribution, two parameters are extracted, i) the annihilation fraction with low momentum electrons, $S$, and ii) that with high momentum electrons, $W$. When positrons annihilate as trapped at open volumes, due to the reduced local electron density, $S$ and lifetime increase, while $W$ decreases depending on the size and nature of the open volumes or defects within the material [13]. For instance, Lhuillier et al. [14,15] and Debelle et al. [16] identified the signal of single vacancies ($V_1$) in W. They monitored its evolution during isochronal annealing experiments and observed that the activation temperature for $V_1$ is between 523-573 K [16].

Theoretically, each type of vacancy defect exhibits specific annihilation characteristics that can be calculated using first principles methods [17,18]. Recently, Yang et al. [19] demonstrated an accurate calculation of the DB Spectrum (DBS) and the positron lifetimes for several transition metals using the Two-components density functional theory (TC-DFT) developed by Makkonen et al. [20]. In tungsten, the evolution of the theoretical DBS for different types of vacancy defects, ranging from $V_1$ to vacancy clusters, agreed with experimental data [21].

PAS can be used to determine defect concentration, with PALS being recognized as the most quantitative technique, as it enables the deconvolution of the exponential components within experimental spectra, allowing for the identification of the nature of defects [14,22,23] and determining their respective concentrations. However, the deconvolution is typically limited to three, or at most four components [23–25], which constrains systematical estimation of the proportion of individual types of defects. In this work, we propose a novel method for quantifying the proportion of a complete distribution of vacancy-type defects. This method integrates the positron trapping model [13,26], taking into account various vacancy defects identified by their theoretical annihilation characteristics, into the simulated annealing (SA) algorithm [27] (see supplementary III b). The vacancy defect distribution was extracted from experimental DBS obtained in self-ion irradiated tungsten at an irradiation dose of 0.0085 and 0.085 dpa at room temperature (RT), and 0.02 dpa at 500 and 700°C. The obtained size distributions are compared to simulations (Molecular Dynamics (MD), Object Kinetic Monte-Carlo (OKMC)) as detailed in Supplementary I. and Transmission Electron Microscopy (TEM) results. The PAS and TEM results are described in Supplementary II.

TEM was used to observe cavities in the irradiated samples. In principle, HR-TEM is sensitive to cavities or vacancy clusters with a size above 0.2 nm, but quantification is affected by the contrast-induced in images by dislocation-type defects (lines or loops). In W, a large density of dislocation loops is formed during self-ion irradiation [28–30] and determination of the cavity concentration with good statistics can only be achieved for cavities larger than 1 nm. Vacancy cluster distributions obtained from MD simulation are extracted from reference [31]. Additionally, OKMC simulations were conducted to model the irradiation-induced evolution of microstructure at higher temperatures, where diffusion of vacancy-type defects is present. These simulations were performed using the MMonCa code [32], and are explained in Supplementary I.

## Vacancy defect distribution for irradiation at RT

Notably, for irradiation at RT with 2 MeV self-ions [33], as explained in supplementary II, the $L_{eff}^+$ and consequently, the total trapping rate $k_{tot}$ determination is less precise. It follows that the positron trap concentration cannot be correctly estimated, only the concentration fraction, independent of the $L_{eff}^+$, can be determined. The proportions of $V_n$ are extracted from the PAS results, combining SA and the trapping model as defined in supplementary III for only the pure V-clusters, giving the PAS-SA $V_n$ distribution. The $V_1$ proportion represents 99% or more of the total concentration of vacancy-type defects at a damage level of 0.0085 dpa. When the damage level increases to 0.085 dpa, the experimental $S$ and $W$ values clearly show the formation of vacancy clusters. The proportion of $V_1$ accounts for 65% (± 32%) of the total concentration, while $V_2$ and $V_3$ fractions are 28% (± 14%) and 5% (± 2%) and larger vacancy clusters remain at a low total proportion of approximately 2 %. The $S$ and $W$ parameters were calculated from the PAS-SA vacancy distributions (see red open squares in Fig.3) and the results overlap the experimental data.

It is commonly accepted that it is very challenging to conduct an accurate quantitative TEM analysis in a damaged W sample due to the influence of dislocation loops on the contrast of the TEM image. This is particularly difficult when irradiation is performed at RT, where the density of loops is high and the size of the vacancy clusters is limited, due to the high migration energy (1.66 eV [34]) of vacancies. Accordingly, the PAS-SA vacancy distribution at RT is only compared to simulation results, as shown in Fig.1 for 0.085 dpa.

At the lowest damage level of approximately 0.0085 dpa, where the overlap of cascades is low, both MD and OKMC simulations indicate that around 95% of vacancy-type defects exist as isolated single vacancies ($V_1$). These results are consistent with PAS-SA results. However, at a higher damage level of 0.085 dpa, MD and OKMC simulations predict higher fractions of isolated $V_1$ compared to PAS-SA (65% ± 32%), with MD showing over 80% and OKMC exceeding 90%. Correspondingly, lower fractions for $V_2$ and $V_3$ are found in simulations (for MD 11% and 2.5% and for OKMC 7% and 1% respectively). These differences could be due to the values of the specific trapping coefficients used to extract the PAS-SA vacancy distributions, which are extrapolated from experimental data (see supplementary II). The $S$ and $W$ parameters were calculated from the MD and OKMC vacancy distributions (see orange, and green open squares in Fig.3) using the trapping model and defined trapping coefficients (see supplementary III). The calculated $S$-$W$ signals show noticeable differences when compared to the experimental ones. The results derived from the OKMC distribution align more closely with the annihilation characteristics of $V_1$. In contrast, the results from the MD distribution are positioned between the annihilation characteristics of $V_2$ and $V_3$. It should be noticed that MD and OKMC simulations were performed using a fixed primary knock-on atom (PKA) energy of 30 keV. According to the Stopping and Range of Ions in Matter (SRIM) calculations, the PKA energy distribution is large, ranging from 55.3 eV to ~1.9 MeV in the case of 2 MeV self-ion irradiation. However, only 6 % of PKAs exceed 30 keV in energy. A large part of PKA has low energy producing smaller defects, such as $V_1$ [35], which can form small clusters ($V_2$, $V_3$) due to their high density. These low-energy PKA could induce a non-negligible fraction of small clusters ($V_2$, $V_3$), as found in PAS-SA distributions. In addition, the difference between MD and OKMC simulations arises from the inclusion of the heat spike effect [36,37] in MD simulations, leading to a slightly higher concentration of clusters equal to or larger than $V_{15}$. Specifically, MD simulations show a larger cluster concentration fraction of ~1.36 %, compared to 0 % in OKMC. This difference is further amplified by the higher specific trapping coefficient of these larger clusters, resulting in an approximately 19 % annihilation fraction of positrons in large clusters. Consequently, both MD and OKMC need refinement, to more accurately predict vacancy defect distributions [38].

Therefore, although there is a slight discrepancy in the exact percentages, the simulation and experimental size distributions exhibit similar trends, particularly those from the MD simulations. According to MD and PAS-SA results, approximately 99 % of the total concentration consists of $V_1$ to $V_4$, with only a small fraction (<1%) of clusters of more than five vacancies. In contrast, OKMC shows a slightly lower fraction of $V_i$ with i = 4-9 compared to the MD and PAS-SA values.

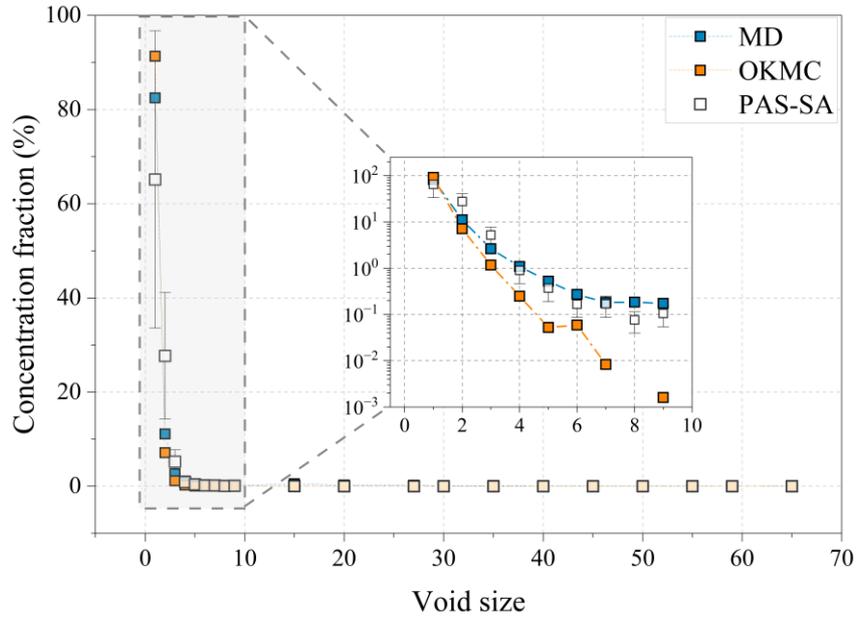

*Fig. 1: Comparison of vacancy concentration fraction obtained from OKMC, and MD simulation [31] for a damage accumulation of about 0.085 dpa compared to a PAS experiment in 2 MeV self-ion irradiated tungsten at RT [33].*

# Vacancy defect distribution for irradiation at high temperatures of 500 and 700 °C

At high temperatures (HT, ≥ 500 °C), vacancies migrate and agglomerate, forming nano-scale clusters, which can be observed using TEM [39,40]. Previous studies combining PAS and TEM [41] have demonstrated the effect of material purity on vacancy aggregation. However, in the previous study, the vacancy cluster size distribution was only determined through TEM observation. Using our new approach by combining the positron trapping model with SA, the vacancy size distribution can be extracted from earlier PAS data [41]. The correspondence between the cavity diameter and the number of vacancies in the cluster was calculated based on the assumption that the volume of a $V_1$ is half of a unit body-centered cubic (BCC) cell of W with a lattice parameter of ~0.315 nm. The volume of a vacancy cluster was then considered proportional to the number of $V_1$ it contains. TEM images showed cavities with sizes ranging from ~0.64 nm (corresponding to $V_8$) to ~1.5 nm ($V_{105}$). The vacancy size distribution, as obtained through TEM and OKMC simulation, was reorganized into twenty distinct subgroups ranging from $V_{20}$ to $V_{105}$ to match with the subgroups chosen as representative of the theoretical annihilation characteristics ($V_1$-$V_{65}$, [21]). Each subgroup represents a specific range of vacancy cluster sizes, with boundaries defined such that the upper and lower limits of each subgroup correspond to the average size between two successive clusters.

Fig.2 compares the size distributions obtained from PAS-SA, TEM, and OKMC for a mean damage accumulation of about 0.02 dpa at 500 °C and 700 °C. For cluster $V_{20}$ or larger, PAS-SA estimates a concentration of the same order of magnitude as that observed by TEM, with PAS-SA values lower by less than a factor of five across all cluster sizes at both temperatures. Notably, the fraction of vacancy clusters larger than $V_{65}$ (1.26 nm in diameter) cannot be extracted from PAS experimental results, because the annihilation characteristics do not change for larger clusters. Clusters larger than $V_{65}$ are observed in TEM images, however, their fraction is small, accounting for about 3.3 % and 7.5 % for 500 °C and 700 °C, respectively. It is worth mentioning that, for the clusters smaller than $V_{20}$, the total concentration estimated by PAS-SA and OKMC is an order of magnitude greater than that of larger vacancy clusters ($V_{>20}$) observed by TEM for 500 °C, and this difference is even more pronounced at 700 °C. This could be due to the detection limit of TEM impeding the observation of the smallest clusters (clusters $V_8$ or smaller with a diameter of approximately 0.5 nm or smaller), which can form during irradiation. To affirm this hypothesis, the annihilation characteristics (*S-W*) corresponding to positron trapping at the defects revealed by TEM, were calculated at both high temperatures. They are shown as magenta squares in Fig.3. The calculated *S-W* points overlap with the ones of vacancy clusters containing 30 vacancies ($V_{30}$) and significantly differ from the experimental PAS data (open black squares in Fig.3) [41]. This result confirms that a notable proportion of small clusters was not detected using TEM. Based on the relative annihilation characteristics calculated using DFT [21], these small clusters are expected to have smaller free volumes than $V_5$ to compensate for the contribution of larger vacancies detected by TEM.

Moreover, upon closer examination of Fig.3, the calculated *S-W* value (red square) using vacancy distribution extracted by SA closely aligns with the experimental result for samples irradiated at RT. However, at HT, a discrepancy remains in the *W* parameter compared to experimental data. The *W* parameter, the fraction of positron annihilated with high momentum electrons, is sensitive to the local chemical environment of positron where it annihilates. As shown in previous studies [7,8,42], LEs such as hydrogen, carbon (C), and oxygen (O) are likely to form stable impurity-vacancy complexes at HT[8]. Consequently, pure vacancy clusters alone could be insufficient for accurately estimating vacancy distribution. Among these impurities, oxygen plays a significant role due to its low migration energy (approximately 0.2 eV [43]) and strong binding affinity with vacancies, as predicted by first-principles calculations [43]. Its contribution appears to be the most pronounced among LEs. Experimental evidence has confirmed the formation of oxygen-vacancy complexes in electron-irradiated tungsten even at RT [7], highlighting the importance of accounting for such impurities in vacancy behavior at elevated temperatures. Secondary Ion Mass Spectroscopy (SIMS) was employed to quantify the concentration of C and O in the irradiated sample. A significant concentration of O was revealed, whereas the C concentration is below the detection limit (as summarized in Supplementary IV),

For this reason, we introduced an additional annihilation state in the SA-trapping model calculation: a single vacancy combined with one oxygen atom ($O_1$-$V_1$). The *S* and *W* values of this complex are estimated from TCDFT results using the same method as for the pure vacancy clusters [7] (see supplementary III). Taking this new defect into account increases the concentration of each vacancy defect by a factor of about 2-3, reaching approximately $10^{23}$ m$^{-3}$ at both 500 °C and 700 °C for the largest V clusters (> $V_{20}$). Their total concentration becomes closer to the TEM results. Consequently, as shown in Fig.3, the corresponding *S-W* values approach experimental values. They coincide with the experimental PAS result at 500 °C, while a minor difference is still observed for 700 °C, which remains within the error bar. It is important to first emphasize that, while TCDFT is the most suitable approach for identifying trends in the variation of $S$ and $W$ values across different vacancy defects, it does not yield absolute values. Consequently, the $S$ and $W$ values for the $O_1$-$V_1$ complex are approximate. On the other hand, despite some intriguing findings obtained through simulations [8], the behavior of $O_m$-$V_n$ complexes at high temperatures remains unclear, and these complexes could potentially contribute to positron trapping, especially at 700°C. Additionally, the possibility that other vacancy complexes may play a role in positron trapping cannot be ruled out. Notably, $O_1$-$V_1$ complexes account for more than 60 % of the total $V_1$ vacancies, with a concentration of approximately 130 at. ppm, which falls within the oxygen concentration range obtained by SIMS (in the Supplementary IV). Furthermore, OKMC predicts a total concentration of small vacancy defects ($<V_{20}$) that closely matches PAS-SA values at 500 °C. However, for irradiation conducted at 700 °C, the OKMC-predicted concentration is over an order of magnitude lower than that of PAS-SA. A possible explanation lies in the influence of the LEs, which can slow down vacancy mobility and form stable vacancy complexes, as predicted in iron (Fe) where single vacancies associated with LEs (e.g., $V_1X_{1-2}$, where X = C, O, or N) exhibit reduced mobility [44]. Given the similarities between W and Fe, a comparable effect could be expected in W. It should be noted that the current OKMC model does not account for LEs, which may explain the discrepancy observed in the concentration of small vacancy defects ($<V_{20}$) between PAS-SA and OKMC, which could potentially be more pronounced at higher temperatures.

Finally, taking into account these small vacancies in the irradiation-induced open volume, swelling was revised to increase significantly to 0.6 ($\pm$ 0.3) % at 500°C, and 0.6 ($\pm$ 0.5) % at 700°C compared to the values estimated from TEM results 0.09 ($\pm$ 0.02) % and 0.04 ($\pm$ 0.02) % respectively [41].

In conclusion, we demonstrated the feasibility of estimating the concentration and the distribution of vacancy clusters at both RT and HT, with results aligning well with computational results and TEM observations, respectively. A subset of vacancy-type defects, invisible to TEM (<$V_{20}$, ~0.85 nm in diameter), was revealed from PAS experimental data. Their concentration is substantial, exceeding $10^{25}$ m$^{-3}$, for irradiation at HT and escalates with the presence of the O impurity. Accounting for these small vacancies increases significantly the irradiation-induced swelling by an order of magnitude. These findings underscore the importance of considering the role of small vacancy clusters and impurity-vacancy complexes, such as oxygen ones, in the evolution of defects in irradiated tungsten.

In addition, the combination of SA and the positron trapping model can be further extended to interpret PAS experiments in other single-elemental materials, such as metals or semiconductors. This approach could more precisely uncover the vacancy-type defects distribution. Such data are significant for discerning a deeper understanding of material behavior under irradiation, in particular regarding swelling and the critical issue of hydrogen trapping in the fusion energy production context.

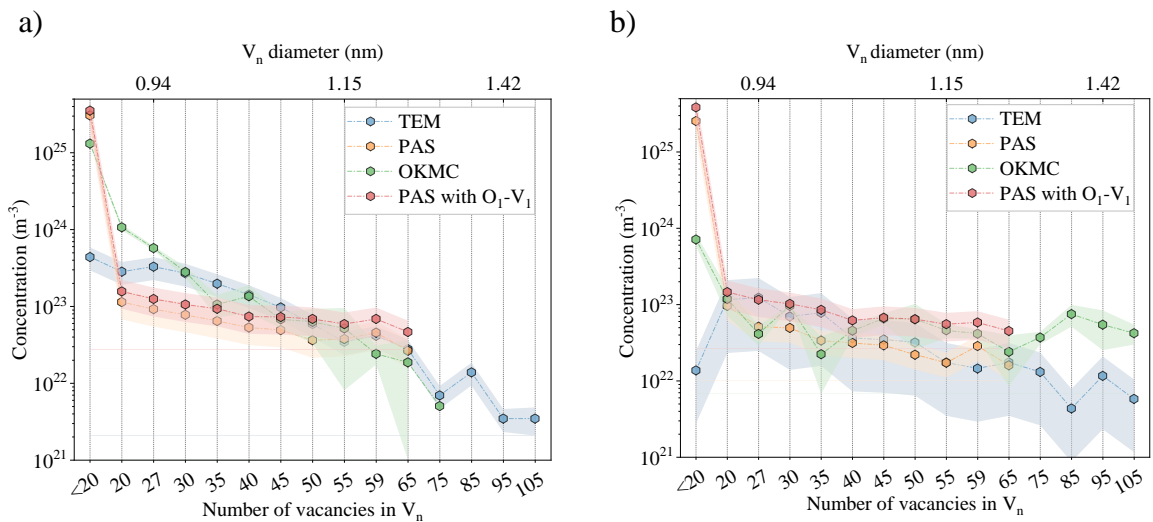

*Fig. 2: Comparison of estimated vacancy concentration obtained from TEM, PAS, and OKMC for a damage accumulation of about 0.02 dpa, a: 500 °C and b: 700 °C [41]. The diameter of the vacancies was estimated using the assumption that the volume of a $V_1$ is half of that of a unit cell of tungsten with a lattice parameter of about 0.315 nm.*

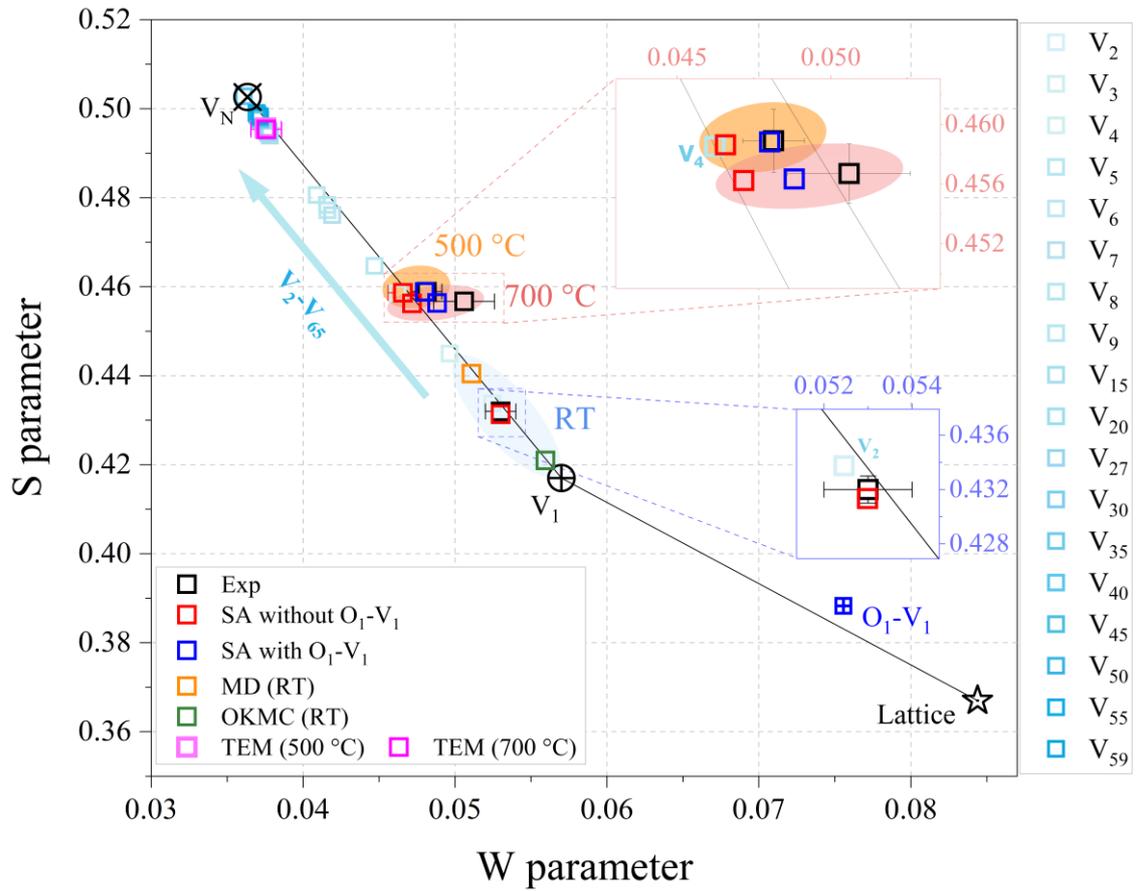

*Fig. 3: Comparison of S-W values estimated from MD, OKMC, TEM and PAS experimental data using two SA models for W self-damaged samples at about 0.085 dpa at RT [33], and 0.02 dpa at 500 °C, 700 °C [41]. The S-W values calculated from the vacancy cluster distribution detected by TEM, or calculated by OKMC, and MD are plotted in magenta, green, and orange, respectively. The theoretical S-W values for the different pure vacancy clusters V1-65 are plotted in degraded blue (intensity increases with the size). The S-W values transposed from DFT calculation for the $O_1$-$V_1$ oxygen-vacancy complex are also plotted as the blue crossed square.*

**Credit authorship contribution statement**

*Zhiwei Hu*: Experimental data processing and interpretation, writing – original draft & editing. *Jintong Wu*: Computational data processing and interpretation, writing – original draft & editing. *François Jomard:* SIMS experiments, data processing, and interpretation *Fredric Granberg:* Conceptualization, supervision, data processing and interpretation, writing – review & editing. *Marie-France Barthe*: Conceptualization, supervision, data processing and interpretation, writing – review & editing.

**Acknowledgments**

This work has been carried out within the framework of the EUROfusion Consortium, funded by the European Union via the Euratom Research and Training Program (Grant Agreement No 101052200 — EUROfusion). Views and opinions expressed are however those of the authors only and do not necessarily reflect those of the European Union or the European Commission. Neither the European Union nor the European Commission can be held responsible for them. This work was partially carried out under the DEVHIS project, funded by the Academy of Finland (Grant number 340538, J.W. and F.G.). Computer time granted by the IT Center for Science -- CSC -- Finland is gratefully acknowledged.

**Declaration of Competing Interest**

The authors declare that they have no known competing financial interests or personal relationships that could have appeared to influence the work reported in this paper.